\documentclass[fleqn]{article}
\date{}
\usepackage[dvips]{graphicx}  
\begin{document}
\title{\bf LOW ENERGY QUARK-GLUON PROCESSES FROM EXPERIMENTAL DATA
USING THE GLOBAL COLOUR MODEL\thanks{Contribution to the
Workshop on Methods of Non-Perturbative Field Theory, Adelaide, Australia 2-13 Feb 1998.}}
\vskip1.0cm
\author{{Reginald T. Cahill  and      Susan M. Gunner
  \thanks{E-mail: Reg.Cahill@flinders.edu.au,
Susan.Gunner@flinders.edu.au}}\\
{  }\\
  {Department of Physics, Flinders University}\\ { GPO Box 2100, Adelaide 5001,
Australia }}

\maketitle

\begin{center}
\begin{minipage}{120mm}
\vskip 0.6in
\begin{center}{\bf Abstract}\end{center}
{The Global Colour Model(GCM) of QCD is used to extract  low energy quark-gluon
processes from experimental data. The resultant effective quark-quark coupling correlator is compared with
that of Jain and Munczek, and with the combined lattice results of Marenzoni {\it et al.} and
Skullerud, and with the two-loop form. The results suggest that higher order  gluon vertices are
playing a role in coupling quark currents. The success of the GCM is explained by an infrared
saturation effect.}\\


\end{minipage} \end{center}

\section{Introduction}

We report on the current status of the project  extracting  low energy quark-gluon processes from
experimental data using the Global Colour Model\cite{CR} (GCM) of QCD. A new GCM  fit (GCM98) to  more
extensive experimental data  is reported here; this updates  previous fits GCM95 and GCM97.  The
effective quark-quark coupling correlator  is compared with that of
Jain and Munczek\cite{JM},  with the combined lattice results of Marenzoni {\it et
al.}\cite{Marenzoni} and Skullerud\cite{Skul98}, and with the two-loop form. The results
show that this   correlator agrees  remarkably well with that of Jain and Munczek, and with one
constructed from the  Marenzoni {\it et al.} and Skullerud   lattice results down to $s=1.8$GeV$^2$. But,
significantly, 
all three of these depart from the two-loop form below $s=2.5$GeV$^2$. The difference between
GCM98 - Jain-Munczek  and the lattice construction below $s=1.8$GeV$^2$ could be an indication of
contributions to the quark-quark coupling at low energy from $n \geq 4$  gluon vertices.

\vspace{5mm}

\section{Global Colour Model}  

The GCM modelling of QCD is based on the idea that as   hadronic correlators are given
by explicit functional integrals  it should be possible, after an appropriate change of
variables of integration, to identify a dominant configuration. It turns out that 
 this  dominant configuration is nothing more than the constituent quark(CQ) effect allowing QCD to be
directly related to  low energy  hadronic physics.   There are a number of reviews of the
GCM\cite{RTC,RW,CG97a,Tandy,
CG98}.    In the functional integral approach correlators are defined by 
\begin{equation} {\cal G}(..,x,...)=\frac{\int{D}\overline{q}{D}q{DA}{D}\overline{C}{D}C....q(x).....
\mbox{exp}(-S_{QCD}[A,\overline{q},q,\overline{C},C])} {\int{D}\overline{q}{D}q{DA}{D}\overline{C}{D}C
\mbox{exp}(-S_{QCD}[A,\overline{q},q,\overline{C},C])}.
\label{eq:2.1}\end{equation}
The  various complete (denoted by scripted symbols) correlators ${\cal G}$ lead to experimental
observables. They  are related by an infinite set of coupled Dyson-Schwinger Equations (DSE), and
by the Slavnov-Taylor gauge-symmetry related identities and,  in the chiral limit, to the axial
Ward-Takahashi identity (AWTI).  The usual truncation of these DSE causes the violation of all
these identities, and in particular of the  AWTI leading to spurious effects for the dynamical breaking of
chiral symmetry which is critical to low energy hadronic physics. One remedy for the AWTI problem is to
modify the quark DSE so that, in conjunction with the meson equations, the AWTI is satisfied.
The GCM avoids this AWTI problem as it  does  not derive from these DSE; rather its nature
follows  from an analytical continuum estimation  procedure for the functional integrations in which the
AWTI is automatically satisfied.  The  correlators in Eq.(\ref{eq:2.1}) may be extracted from the
generating functional of QCD, 
\begin{equation}
Z_{QCD}[\overline{\eta},\eta,J]=\int {D}\overline{q}{D}q{D}A{D}\overline{C}{D}C\mbox{exp}(-S_{QCD}[A,
\overline{q},q,\overline{C},C]+\overline{\eta}q+
\overline{q}\eta+JA).
 \label{eq:2.2b}\end{equation}
Functional transformations lead\cite{RTC} to the GCM; briefly  
and not showing source terms for convenience,  the gluon and ghost integrations are formally
performed
\begin{eqnarray*}\int {D}\overline{q}{D}q{D}A{D}\overline{C}{D}C\mbox{exp}(-S_{QCD}) = \int {D}\overline{q}{D}q
\mbox{exp}(-\int 
\overline{q}(-\gamma . \partial+{\cal M})q +   \end{eqnarray*}
\begin{equation}\mbox{\ \ \ \ \ \ \ \ \ \ \ \ \ \ \ \ \ \ \ \ \ \ \ }
+\frac{g_0^2}{2}\int
j^a_{\mu}(x)j^a_{\nu}(y){\cal G}_{\mu\nu}(x-y)+\frac{g_0^3}{3!}\int
j^a_{\mu}j^b_{\nu}j^c_{\rho}{\cal
G}^{abc}_{\mu\nu\rho}+......),\label{eq:2.4}\end{equation}where
$j^a_{\mu}(x)=\overline{q}(x)\frac{\lambda^a}{2}\gamma_{\mu}q(x)$, $g_0$ is the bare coupling
constant, and ${\cal G}_{\mu\nu}(x)$ is  the  gluon
correlator  with no quark loops but including ghosts ($\overline{C},C$)
\begin{equation}{\cal G}_{\mu\nu}(x-y)=
\frac{\int {D}A{D}\overline{C}{D}CA^a_{\mu}(x)A^a_{\nu}(y)\mbox{exp}(-S_{QCD}[A,\overline{C},C])}
{\int {D}A{D}\overline{C}{D}C\mbox{exp}(-S_{QCD}[A,\overline{C},C])}.
\label{eq:2.5}\end{equation}  
 Fig.1 shows successive terms in Eq.(\ref{eq:2.4}). The terms of higher order than
the term quartic in the quark fields   are difficult to explicitly  retain in any analysis. 
However the GCM models the effect of  higher order terms by replacing the
coupling constant
$g_0$ by a momentum dependent quark-gluon coupling $g(s)$, and neglecting terms like ${\cal
G}^{abc}_{\mu\nu\rho}$  and higher order in Eq.(\ref{eq:2.4}). This $g(s)$ is a restricted form of vertex
function.  The modification 
$g_0^2{\cal G}_{\mu\nu}(p) \rightarrow D_{\mu\nu}(p) = g(p^2){\cal G}_{\mu\nu}(p)g(p^2)$
and the truncation  then defines the GCM.  We also call this effective quark-quark coupling
correlator $D_{\mu\nu}(p)$ the effective gluon correlator.

\begin{minipage}[t]{35mm}
\hspace{5mm}\includegraphics[scale=0.3]{Fig1a.EPSF}

 \makebox[20mm][c]{(a)}
\end{minipage}
\begin{minipage}[t]{40mm}  
\hspace{5mm}\includegraphics[scale=0.3]{Fig1b.EPSF}
\makebox[28mm][c]{(b)}
\end{minipage}
\begin{minipage}[t]{40mm}
 \hspace{5mm}\includegraphics[scale=0.3]{Fig1c.EPSF}
\makebox[20mm][c]{(c)}
\end{minipage}
\begin{figure}[ht]
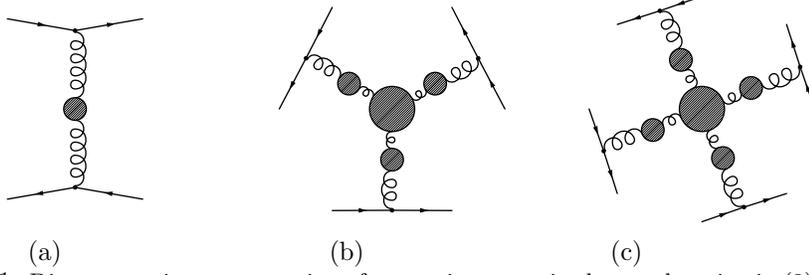

\vspace{-7mm}
\caption{\small{Diagrammatic representation of successive terms in the quark action in 
(\ref{eq:2.4}).  The quark-gluon vertex strength is $g_0$, while the
gluon-gluon vertices (including gluon correlators)  are fully dressed except for
quark loops.}
 \label{figure:QCDaction}}
\end{figure}

\begin{minipage}[t]{35mm}
\hspace{5mm}\includegraphics[scale=0.3]{Fig2a.EPSF} 
 \makebox[20mm][c]{(a)}
\end{minipage}
\begin{minipage}[t]{40mm}
\vspace{-30mm}  
\hspace{0mm}\includegraphics[scale=0.4, bb =0 +40 400 250]{Fig2b.EPSF}
\makebox[28mm][c]{(b)}
\end{minipage}
\begin{minipage}[t]{40mm}
 \hspace{5mm}\includegraphics[scale=0.3]{Fig2c.EPSF}
\makebox[20mm][c]{(c)}
\end{minipage}
\begin{figure}[ht]
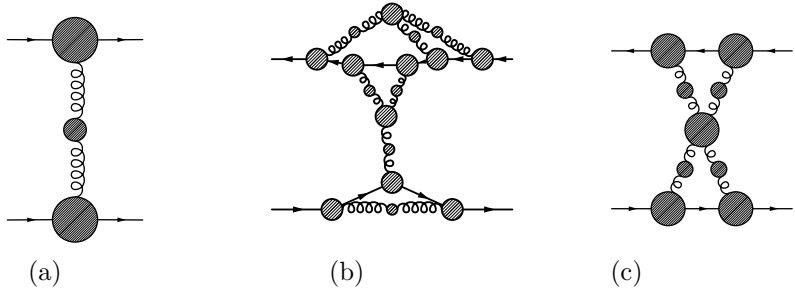

\vspace{-5mm}\caption{\small{(a) The GCM effective $D_{\mu\nu}$ in
(\ref{eq:2.6}), (b) example of correlations formally included  in (a), and in (c)  an $n=4$ process
not formally included in (a), but which is modelled in the GCM  via the specific form of
$D_{\mu\nu}$.}
\label{figure:Ints} }
\end{figure}

\begin{figure}[ht]
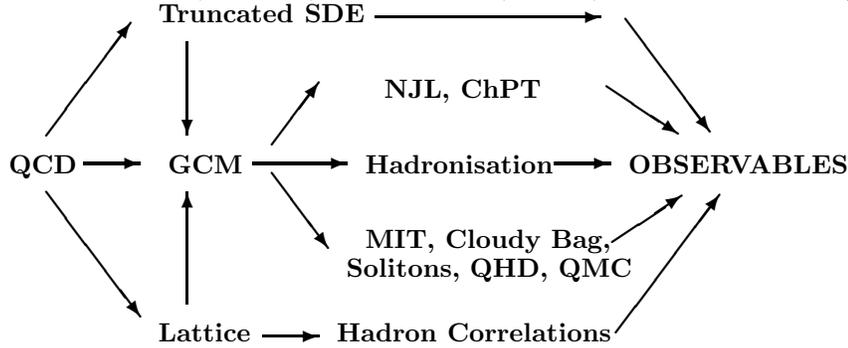
 
\vspace{-15mm}\caption{\small{Relational map of the GCM to QCD and various other
modellings including the Nambu - Jona-Lasinio (NJL), Quantum Hadrodynamics (QHD),
Quark Meson Coupling (QMC)  and Chiral Perturbation Theory (ChPT).}\label{figure:Map}} 
\end{figure}

  The GCM is equivalent to using a quark-gluon field theory with the action
\begin{equation} S_{GCM}[A,\overline{q},q]=\int \left( 
\overline{q}(-\gamma . \partial+{\cal M}+iA^a_{\mu}\frac{\lambda^a}{2}\gamma_{\mu})q
 +\frac{1}{2} A^a_{\mu}D^{-1}_{\mu\nu}(i\partial)A^a_{\nu} \right).\label{eq:2.6}
\end{equation} 
Here $D_{\mu\nu}^{-1}(p)$ is the matrix inverse of the Fourier transform of $ D_{\mu\nu}(x)$ and Fig.2
shows processes formally included in $D_{\mu\nu}(p).$  This action is invariant under $q\rightarrow Uq,
\overline{q}\rightarrow \overline{q}U^\dagger$, and $A^a_{\mu}\lambda^a \rightarrow U
A^a_{\mu}\lambda^a U^\dagger$  (where $U$ is a global $3\times3$ unitary colour matrix) - the
global colour symmetry of the GCM.
 The gluon  self-interactions  that
arise as a consequence of  the local colour symmetry in Eq.(\ref{eq:2.5}) and the ghost and vertex
effects   lead to 
$D_{\mu\nu}^{-1}(p)$ being  non-quadratic. Hence, in effect, the GCM models the QCD local
gluonic action  $\int F^a_{\mu\nu}[A]F^a_{\mu\nu}[A]$ in
$S_{QCD}$ of Eq.(\ref{eq:2.1}) which has local colour symmetry, by a highly
nonlocal action in the last term of Eq.(\ref{eq:2.6}) which has global colour symmetry. It is
important to appreciate that while the GCM has a formal global colour symmetry the detailed dynamical
consequences of the local colour symmetry of QCD are modelled by the detailed form of $D(s)$. 
Approximations to the truncated DSE usually map QCD onto the GCM, as indicated in Fig.\ref{figure:Map}.
The success of this GCM modelling has been amply demonstrated
\cite{RTC,RW,CG97a,Tandy,CG95}.

Hadronisation\cite{RTC} of the functional integrations in Eq.(\ref{eq:2.1})
 involves a sequence of  changes of
variables involving, in part, the transformation to bilocal boson fields, and
then to the usual local hadron fields (sources not shown): 
\begin{eqnarray*}  Z\approx\int {D}\overline{q}{D}q{D}A\mbox{exp}(-S_{GCM}[A,\overline{q},q]+\overline{\eta}q+
\overline{q}\eta)   \mbox{\ \ \ \ (GCM)}
\end{eqnarray*}  
\begin{equation}
\mbox{\ \ } =\int {D}{\cal  B}{D}\overline{D}{D}D\mbox{exp}(-S[
{\cal B},\overline{\cal D},{\cal D}]) \mbox{\
\ \ \ (bilocal fields)}
\label{eq:2.7}\end{equation} 
\begin{equation} \mbox{\ \ } =\int{D}\overline{N}{D}N..{D}\pi{D}\rho{D}\omega...\mbox{exp}(-S_{had}[\overline{N},N,..\pi,\rho,\omega....]) 
  \mbox{\ \ \ \ (local fields) }.\label{eq:2.8}\end{equation} 
The bilocal fields in Eq.(\ref{eq:2.7}) arise  naturally and correspond to the fact that, for
instance, mesons are extended states.  This
hadronisation  derives from functional integral calculus changes of
variables which are induced by generalised Fierz transformations that emerge from the colour,
spin and flavour structure of QCD.
The final functional integrations in Eq.(\ref{eq:2.8}) over the hadrons  give the
hadronic observables, and are equivalent to dressing each constituent hadron by, mainly, lighter
constituent mesons.    The basic insight is that the quark-gluon dynamics, in Eq.(\ref{eq:2.1}),  is
fluctuation dominated, whereas the hadronic functional integrations in  Eq.(\ref{eq:2.8}) are
not.  The induced  hadronic effective action in Eq.(\ref{eq:2.8}) is nonlocal.  The GCM automatically
preserves the consequences of the dynamically broken chiral symmetry in the action in Eq.(\ref{eq:2.8}). 
\normalsize

\vspace{5mm}
\setlength{\unitlength}{0.25mm}
\begin{center}
\hspace{-30mm}
\begin{picture}(10,180)(+200,80)
\thicklines
\put(5,200){\bf QCD}
\put(25,190){\vector(3,-4){50}}
\put(45,205){\vector(3,0){30}}
\put(85,200){ \bf GCM}
\put(145,215){\vector(3,4){25}}
\put(200,240){ \bf NJL, ChPT}
\put(135,205){\vector(3,0){50}}
\put(190,200){ \bf Hadronisation}
\put(295,205){\vector(3,0){30}}

\put(333,282){\vector(3,-4){45}}
\put(323,246){\vector(3,-2){38}}
\put(326,163){\vector(3,+2){37}}
\put(328,115){\vector(3,+4){55}}

\put(330,200){ \bf OBSERVABLES}
\put(145,200){\vector(3,-4){30}}
\put(190,160){ \bf MIT, Cloudy Bag,  }
\put(180,145){ \bf Solitons, QHD, QMC}
\put(80,110){ \bf Lattice }
\put(140,113){\vector(1,0){30}}
\put(100,130){\vector(0,1){60}}
\put(175,110){ \bf Hadron Correlations}
\put(80,280){ \bf Truncated SDE }
\put(200,283){\vector(1,0){120}}
\put(25,220){\vector(3,+4){45}}
\put(100,270){\vector(0,-1){50}}

\end{picture}
\end{center}

A key idea in the GCM is that in proceeding from  Eq.(\ref{eq:2.7}) to Eq.(\ref{eq:2.8})
one expands $S[{\cal B},..]$    about the configuration ${\cal
B}_{CQ}$ that  minimises it, giving in Eq.(\ref{eq:CQ})  the GCM Constituent Quark (CQ)
equations,
\begin{equation}
\frac{\delta S}{\delta {\cal B}(x,y)}\left|_{{\cal B}_{CQ}}=0\right.. 
\label{eq:Min}\end{equation}
 Thus for 
{\it all} hadrons one assumes a universal dominant configuration.  This amounts to assuming
that all hadrons share a  common dynamical feature.
Of the set ${\cal B}(x,y)_{CQ}$ only  $A(x-y)$ and $B(x-y)$ (their Fourier transforms appear in
Eq.(\ref{eq:G})) are non-zero translation-invariant bilocal fields characterising the dominant
configuration. This is the dynamical breaking of chiral symmetry. The dominant configuration is
analogous to the condensate  of Cooper pairs in the BCS theory of superconductivity. However unlike
solid state  superconductivity here there is no normal component.
 Writing out the translation invariant CQ equations
 we find that the dominant configuration is indeed simply the constituent  quark
effect as they may be written  in the form, 
\begin{equation}
G^{-1}(p)=i\backslash \!\!\!p +m+
\frac{4}{3}\int\frac{d^4q}{(2\pi)^4}D_{\mu\nu}(p-q)\gamma_{\mu}G(q)\gamma_{\nu},
\label{eq:CQ}\end{equation}
and we see that this is  the gluon dressing of a constituent quark. This equation is  {\it exact}
in the GCM and its solution has the structure  
\begin{equation}
G(q)=(iA(q)q.\gamma+B(q)+m)^{-1}=-iq.\gamma\sigma_v(q)+\sigma_s(q).
\label{eq:G}\end{equation}
 In the chiral limit there are more ${\cal
B}_{CQ}$  fields that are non-zero, and the resultant degeneracy of the dominant
configuration is responsible for the masslessness of the pion.  
The  constituent quark  correlator  $G$ should not be confused with the 
 complete quark correlator ${\cal G}$ from Eq.(\ref{eq:2.1}).    This ${\cal G}$ would be needed
to analyse the existence or otherwise of free quarks. The $G$ on the other hand relates
exclusively to the internal structure of hadrons, and to the fact that this appears to be
dominated by the constituent quark effect. The evaluation of  ${\cal G}$ is a very difficult
task, even within the GCM while $G$ is  reasonably easy to study using Eq.(\ref{eq:CQ}). The 
truncation of the DSE in which the full quark ${\cal G}$ is approximated by the CQ $G$ amounts to
using a mean field approximation; however the truncated DSE then has no systematic formalism  for
going beyond the mean field as in the GCM.
 The hadronic effective action in Eq.(\ref{eq:2.8}) arises when $S[{\cal B},..]$ is expanded
about the dominant CQ configuration; the first derivative is zero  by Eq.(\ref{eq:Min}), and the
second derivatives, or curvatures, give the constituent or core meson correlators $G_m(q,p;P)$
\begin{equation}
 G_m^{-1}(q,p;P)=F.T.\left(\frac{\delta^2 S}{\delta {\cal B}(x,y)\delta {\cal
B}(u,v)}\left|_{{\cal B}_{CQ}}\right.\right),
\label{eq:CM}\end{equation}
after exploiting  translation invariance and Fourier transforming. Higher order
derivatives  lead to couplings between the  meson cores.  The $G_m(q,p;P)$ are given by 
ladder-type  correlator equations. The non-ladder effects are inserted by
the final functional integrals in  Eq.(\ref{eq:2.8}), giving the complete GCM meson correlators
${\cal G}_m(q,p;P)$.  It is interesting
to note that the truncated and modified DSE in Maris and Roberts\cite{MR} are identical to
 Eq.(\ref{eq:CQ}) and Eq.(\ref{eq:CM}) (in the form of Eq.(\ref{eq:meson})). However Maris and
Roberts\cite{MR} assume that $D(s)$ has  the two loop form down to $s\approx 1$GeV$^2$, and this
assumption is not supported by the detailed analyses reported herein.  An important advantage of the GCM
modelling of QCD is that the hadronic sector  effective action is manifestly derivable. One key feature
is that the meson exchange within the diquark subcorrelations  of the baryon correlators is seen to insert 
crossed   gluonic exchanges, and these appear to stop the diquark correlations from developing a mass-shell,
i.e. the diquark correlations are confined\cite{BRvS}.

 In the present analysis the $\omega$ and a$_1$ mesons are described by
these constituent meson correlators; that is, we ignore meson dressings of these mesons. 
The mass $M$ of these states is determined by finding the pole position of $G_m(q,p;P)$ in the
meson momentum $P^2=-M^2$ and this leads to the homogeneous vertex equation
\begin{equation}\Gamma(p;P)=-\frac{4}{3}\int\frac{d^4q}{(2\pi)^4}D_{\mu\nu}(q-p)
\gamma_{\mu}G(q+\frac{P}{2})\Gamma(q;P)G(q-\frac{P}{2})\gamma_{\nu}\label{eq:meson}.
\end{equation} 

The success of the GCM appears to be based on the phenomenon of an IR saturation
mechanism in which the extreme IR strength  of the many contributing  quark-quark
couplings  is easily modelled by this one effective gluon correlator  $D_{\mu\nu}(p)$.    Of particular
dynamical interest  is the comparison  of the GCM $D_{\mu\nu}(p)$ with one constructed theoretically
from only a  gluon correlator and vertex functions, say from continuum or lattice modellings, for this  
gives some insight into the IR strength of the higher order gluonic couplings.

\section{Procedures}

 To solve Eq.(\ref{eq:CQ}) for various $D_{\mu\nu}(p)$ and then to proceed to use 
$A(s)$ and $B(s)$ in meson correlator equations for fitting observables to meson data is
particularly difficult. A  robust numerical technique is to use a separable
expansion\cite{CG95}. 
\begin{equation}
D_{\mu\nu}(p)=(\delta_{\mu\nu}-\frac{p_{\mu}p_{\nu}}{p^2})D(p^2), \mbox{\ \ and \ \ } 
{\cal G}_{\mu\nu}(p)=(\delta_{\mu\nu}-\frac{p_{\mu}p_{\nu}}{p^2}){\cal G}(p^2).
\label{eq:LG}\end{equation}
We expand   $D(p-q)$ in Eq.(\ref{eq:CQ}) into $O(4)$ hyperspherical harmonics
\begin{equation}
D(p-q)=D_0(p^2,q^2)+q.pD_1(p^2,q^2)+...,
\end{equation}
\begin{equation}
D_0(p^2,q^2)=\frac{2}{\pi}\int_0^{\pi}d\beta \mbox{sin}^2 \beta D(p^2+q^2-2pq\mbox{cos}
\beta),...
\label{eq:D}\end{equation}
We then introduce multi-rank separable $D_0$ expansion (here $n=3$):
\begin{equation}
D_0(p^2,q^2)=\sum_{i=1,n} \Gamma_i(p^2)\Gamma_i(q^2).
\label{eq:Gammas}\end{equation}
The constituent quark equations then have solutions of the form
\begin{equation}
B(s)=\sum B_i(s), \mbox{\ \  } B_i(s)=b_i\Gamma_i(s), \mbox{\ \  }\sigma_s(s)=\sum_{i=1,n}\sigma_s(s)_i, 
\label{eq:Bs}\end{equation}
\begin{equation}
 b_i^2=4\pi^2\int_0^{\infty} sds B_i(s)\sigma_s(s),
\mbox{\ \ \  and \ \ }
B_i(s)=\frac{\sigma_s(s)_i}{s\sigma_v(s)^2+\sigma_s(s)^2},
\label{eq:be}\end{equation}

However rather than specifying $\Gamma_i$ in (\ref{eq:Gammas}) we  proceed by
parametrising forms  for the $\sigma_{si}$ and $\sigma_v$; the  $\Gamma_i$
then follow from (\ref{eq:Bs}) and (\ref{eq:be}): 
\begin{eqnarray*}
\sigma_s(s)_1=c_1\mbox{exp}(-d_1s), \mbox{\ \ \ \ }
\sigma_s(s)_2=c_2.\left(\frac{2s^2-d_2(1-\exp(-2s^2/d_2))}{2s^4}\right)^2,
\end{eqnarray*}
\begin{eqnarray*}
\sigma_s(s)_3=c_3\left(\frac{2f(s)-d_3(1-\exp(-2f(s)/d_3))}{2f(s)^2}\right)^2,\mbox{\ }
f(s)=s(ln(\tau +s/\Lambda^2))^{1/2},
\end{eqnarray*}
\begin{equation}                                                          \label{eqn:seps}
\sigma_v(s)=  \frac{2s-\beta^2(1-\exp(-2s/\beta^2))}{2s^2}.
\end{equation}
 The three
$\sigma_{si}$ terms mainly determine the IR,  midrange and UV regions; the $\sigma_s(s)_3$ term
describes the  asymptotic form of $\sigma_s(s)\sim 1/s^2\ln(s/\Lambda^2)$ for $s\rightarrow \infty$ 
 and ensures the  form for $D(s)\sim 1/s\ln(s/\Lambda^2)$.
  With these
parametrised forms  we can numerically  relate the mass of the
$a_1$(1230MeV) and $\omega$(783MeV)  mesons from Eq.(\ref{eq:meson}), 
$f_{\pi}$(93.3MeV)  and  experimental  points for $\alpha(s)$ (see Fig.4a insert) from the Particle Data
Book for
$s >3$GeV$^2$ to the   parameter set in Table 1 in a robust and stable manner.

\vspace{3mm}
\noindent Table 1: $\sigma_s(s)$ and $\sigma_v(s)$  Parameters.
\vspace{2mm}

\hspace{4mm}
\begin{tabular}{|llllll|}
\hline 
$\mbox{ \ \ }$$c_1$ &1.839GeV$^{-1}$  & $d_1$    & 3.620GeV$^{-2}$&$\mbox{ \ \ }\beta$  & 0.4956GeV
\\
$\mbox{ \ \ }$$c_2$ &0.0281GeV$^{7}$  & $d_2$    & 1.516GeV$^{4}$&$\mbox{ \ \ }\Lambda$ &0.234GeV\\
$\mbox{ \ \ }$$c_3$  & 0.0565GeV$^{3}$  &$d_3$  & 0.7911GeV$^2$ & & \\  
\hline
\end{tabular}

\begin{figure}[ht]
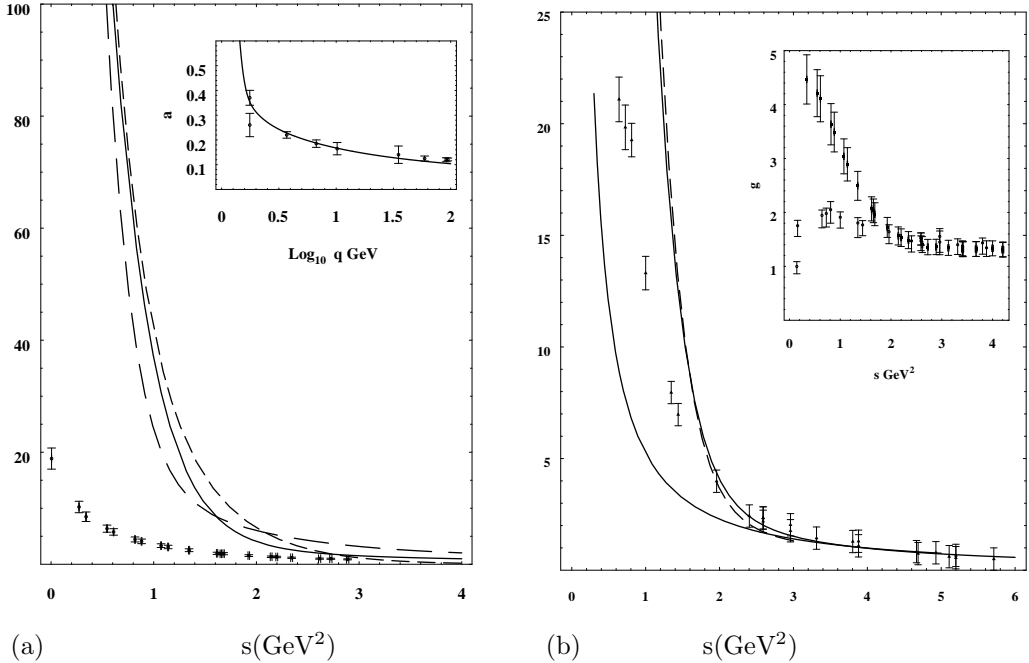

\vspace{-6mm}\caption{Plots of $D(s)$ (GeV$^{-2}$). In (a) solid line is GCM98; shortdash line
is GCM95; longdash line is GCM97. Data plot is lattice pure-gluon ${\cal G}(s)$ from Marenzoni {\it et
al.}, and so has no quark-gluon vertex. Insert is fit of GCM98 to $\alpha(s)$ of Particle Data Book. In (b)
GCM98 is the solid line; dashed line is Jain and Munczek; lower solid line is two-loop form with
$\Lambda=0.234$GeV and $N_f=3$; data plot is combined lattice data for $g(s){\cal G}(s)g(s)$ with ${\cal
G}(s)$ from Marenzoni {\it et al.} (as in (a)) and $g(s)$ from Skullerud. Insert shows $g(s)$ from
Skullerud (lower data plot), and from GCM98/${\cal G}(s)$ (upper data plot).
\label{figure:GluonCorrelators}}
\end{figure}

\vspace{3mm} 
The translation invariant form for the
effective gluon correlator is  easily reconstructed by using $D(p^2)=D_0(p^2,0)$ from
Eq.(\ref{eq:D})
\begin{equation}
D(p^2)=\sum_i\frac{1}{b_i^2}\frac{\sigma_s(0)_i}{\sigma_s(0)^2}\frac{\sigma_s(p^2)_i}
 {p^2\sigma_v(p^2)^2+\sigma_s(p^2)^2}.
\end{equation}
 With the parameter set in Table 1 the resulting $D(p^2)$ is shown in
Fig.\ref{figure:GluonCorrelators}.  
   Shown in Fig.\ref{figure:GluonCorrelators}a is the pure gluon correlator  ${\cal G}(p^2)$ from
the lattice calculations of Marenzoni {\it et al} \cite{Marenzoni}, corresponding to the value
$\beta=6.0$,  where the errors arise mainly from a $5\%$
uncertainty in the lattice spacing; $a=0.50 \pm 0.025GeV^{-1}$. A significant feature of QCD is
that the infrared dominance, as revealed by the large value of $D(s)$ at small $s$, causes the
CQ equations to saturate, i.e. the form of the solutions $A(s)$ and $B(s)$ is independent of the
detailed IR form of $D(s)$.  This saturation effect means  that low energy QCD is
surprisingly easy to model.

\vspace{-1mm}


\hspace{-8mm}\begin{minipage}[t]{70mm}
\includegraphics[scale=1.0]{Fig4a.EPSF}
\makebox[60mm][l]{\mbox{\ \ \ \ }(a) \hspace{25mm}s(GeV$^2$) } 
\end{minipage}
\begin{minipage}[t]{60mm}
\includegraphics[scale=0.84]{Fig4b.EPSF}
\makebox[60mm][l]{\mbox{\ \ \ \ }(b) \hspace{15mm}s(GeV$^2$) }
\end{minipage}

\noindent Also shown in
Fig.\ref{figure:GluonCorrelators}(a)  are the earlier GCM fits: GCM95\cite{CG95} and GCM97\cite{CG97b}.
The insert in Fig.\ref{figure:GluonCorrelators}(a) shows the GCM98 in the form of  $\alpha(q)$
($D(s)=\frac{4\pi\alpha(s)}{s}$) compared with experimental data from  the Particle Data Book. In
Fig.\ref{figure:GluonCorrelators}(b) we show the new GCM98 quark-quark coupling correlator
$D(s)$  showing excellent agreement with the Jain-Munczek\cite{JM} $D(s)$, and with one
constructed from the  Marenzoni {\it et al.} and Skullerud   lattice results down to $s=1.8$GeV$^2$;
$D(s)=g(s){\cal G}(s)g(s)$. The normalisation and shape of the Marenzoni {\it et al.} ${\cal
G}(s)$  agrees with that of Leinweber {\it et al.}\cite{LSW}.
However  the normalisation of the much more difficult lattice computation of  $g(s)$  is uncertain and
we have chosen it  so that the combined lattice $D(s)$ agrees with the experimental Particle  Data Book
for $s>3$GeV$^2$. As shown  in Fig.\ref{figure:GluonCorrelators}(b)  all three of these depart from the
two-loop form below  $s=2.5$GeV$^2$ . 
The difference between GCM98 - Jain-Munczek and the lattice construction could be an
indication of contributions to the quark-quark coupling  from
$n \geq 4$  gluon self-couplings at low energy, since processes like  Fig.\ref{figure:Ints}(c) would be
included in the GCM fit, but are not in the lattice construction.

The insert in Fig.\ref{figure:GluonCorrelators}(b) shows the $g(s)$ from 
$g^2(s)=D(s)/{\cal G}(s)$  that then follows from our analysis. This is the
effective quark-gluon coupling vertex   if the gluon correlator is taken to be that of 
Marenzoni {\it et al.} or Leinweber {\it et al.}. Here the error bars now indicate combined
errors and uncertainties from the lattice spacing.  Also shown is
$g(s)$ from Skullerud\cite{Skul98} with the normalisation as discussed above.  

The GCM95-GCM98 parametrisations differed mainly
in their asymptotic forms, and when fitting to meson data these differences only resulted in 
slight variations of
$D(s)$ for intermediate $s$ values, as shown in Fig.\ref{figure:GluonCorrelators}(a).  The
consequent small variations in the predicted values of various hadronic properties are shown in Table 2.
However forcing the asymptotic form to fit experimental data for $D(s)$  even for
$s>3$GeV$^2$  results in a stabilisation of the GCM $D(s)$.

\vspace{2mm}
\noindent Table 2:  Hadronic Observables.

\vspace{2mm}
\small{
\hspace{-7mm} \begin{tabular}{| l r r r r |} 
\hline 
 {\bf Observable} & {\bf GCM1995}& {\bf GCM1997}& {\bf GCM1998} & {\bf
Expt./Theory} \\
\hline  
$f_{\pi}$  &    93.0MeV*& 93.2MeV*&  92.40MeV*  &93.3MeV\\
$a_1$ meson mass  & 1230MeV* &1231MeV* & 1239MeV* &1230MeV\\
$\pi$ meson mass(for m$_{u,d}$)   &  138.5MeV*   & 138.5MeV* &  138.5MeV*  &138.5MeV\\
     $\alpha(q)$ &- & - &see fig.4a & $\dagger$\\
$K $ meson mass (for $m_s$ only)   &  496MeV*      & -  &-&     496MeV\\
$(m_u+m_d)/2|_R(\mu=1$GeV)   &  6.5MeV & 4.8MeV  &7.7MeV &  $\approx$8.0MeV\\
$m_s|_R(\mu=1$GeV)   &  135MeV   & - &  - &130MeV\\
$\omega$ meson mass &  804MeV  & 783MeV* & 783MeV*  &782MeV  \\
$a^0_0$ $\pi-\pi$ scatt. length &  0.1634  & 0.1622 &  0.1657 &   0.26 $\pm$ 0.05 \\
$a^2_0$ $\pi-\pi$ scatt. length&  -0.0466  &-0.0463 & -0.0465  & -0.028 $\pm$ 0.012 \\
$a^1_1$  $\pi-\pi$ scatt. length&  0.0358   & 0.0355 & 0.0357  &0.038 $\pm$ 0.002 \\
$a^0_2$ $\pi-\pi$ scatt.length&  0.0017   & 0.0016& 0.0018  &0.0017 $\pm$ 0.003\\
$a^2_2$ $\pi-\pi$ scatt.length &  -0.0005  &-0.0005 & -0.0003  &0.00013$\pm$0.0003\\
$r_{\pi}$ pion charge radius & 0.55fm & 0.53fm & 0.53fm  &0.66fm\\
$\frac{1}{2}^+(0^+)$nucleon-core mass$^{**}$  &  1390MeV & 1435MeV & 1450MeV 
&$>$1300MeV$\dagger\dagger$ 
\\ constituent quark rms size   & 0.51fm&0.39fm & 0.58fm  &-\\
chiral quark constituent mass &   270MeV& 267MeV  & 325MeV  &-\\
$0^+$ diquark rms size   & 0.55fm&0.55fm & 0.59fm  &-\\
$0^+$ diquark const. mass &  692MeV  & 698MeV & 673MeV  &$>$400MeV  \\
$1^+$ diquark const.  mass &  1022MeV  &903MeV & 933MeV  & -  \\ 
$0^-$ diquark const.  mass &  1079MeV  & 1049MeV & 1072MeV  &-  \\
$1^-$ diquark const.  mass &  1369MeV  & 1340MeV & 1373MeV  &-  \\
MIT bag constant  &  (154MeV)$^4$  &(145MeV)$^4$ & (166MeV)$^4$  &(146MeV)$^4$ \\
MIT N-core (no cms corr.)   & 1500MeV & 1420MeV  & 1625MeV  &$>$1300MeV$\dagger\dagger$   \\
\hline 
\multicolumn{5}{|l|}{* fitted observable;  -  not computed or not known; $\dagger$
$\alpha(s)$ from Particle Data Book;}\\
\multicolumn{5}{|l|}{GCM1995: Cahill and Gunner\cite{CG95};  GCM1997:  Cahill and
Gunner\cite{CG97b}; GCM1998: this report.}\\
\multicolumn{5}{|l|}{$^{**}$ only $0^+$ diquark correlation; $1^+$ diquark correlation
lowers nucleon core mass.}\\
\multicolumn{5}{|l|}{ $\dagger\dagger$ nucleon core mass (i.e. no meson dressing). }\\
\hline
\end{tabular}}
\normalsize

\vspace{2mm}

\section{Conclusion}

Figs.\ref{figure:GluonCorrelators}(a) and \ref{figure:GluonCorrelators}(b)  reveal that the
GCM95-98 project of extracting low energy quark-gluon processes has reached a stage where some detailed
insights are emerging.  It is clear that the  continuum modelling,  when fitted to experimental
data, leads to a unique quark-quark coupling correlator. This is shown by the excellent
agreement between the GCM multi-rank  technique and the Jain-Munczek result\cite{JM,KK}.  This continuum
result is also in excellent agreement with the combined lattice prediction down to $s \approx
1.8$GeV$^2$. All three results show a clear departure from the two-loop form below
$s=2.5$GeV$^2$. The difference between the GCM modelling and the combined lattice  $D(s)$ at
lower $s$ values raises an interesting interpretation. Assuming that the lattice calculation of
$g(s)$ is confirmed  by future studies, particularly at low $s$,  we must conclude that the
GCM analysis of the experimental data is revealing the influence of processes that are not in the
lattice construct. The most obvious possibility is that the high order gluon vertices, as in 
Fig.\ref{figure:Ints}(c), are contributing at low energies, and providing additional attractive
interaction  between quark currents. Provided these contribute only in the region where $D(s)$
is large the detailed form of these contributions is not needed since the saturation effect
means that hadronic observables are independent of these details, including their gauge dependence.
Thus the saturation of the constituent quark processes and the related dynamical breaking of chiral
symmetry explains why the GCM  works so well as a low-energy {\it equivalent} field theory for QCD.

\newpage

\end{document}